\documentstyle[aps,epsfig,epic,eepic,twocolumn]{revtex}
\draft

\newcommand{\pathfigs}{.}

\newcommand{\be}{\begin{equation}}
\newcommand{\ee}{\end{equation}}
\newcommand{\bea}{\begin{eqnarray}}
\newcommand{\eea}{\end{eqnarray}}
\newcommand{\bdm}{\begin{displaymath}}
\newcommand{\edm}{\end{displaymath}}

\newcommand{\bi}{\begin{itemize}}
\newcommand{\ei}{\end{itemize}}

\newcommand{\refkl}[1]{(\ref{#1})}

\newcommand{\erw}[1]{\mbox{$\langle #1 \rangle$}} 

\newcommand{\rhomax}{\rho_{\rm max}}

\begin{document}
\pagestyle{myheadings}
\markboth{Treiber/Helbing: Macroscopic Simulation of Widely Scattered
Synchronized Traffic States}
{Treiber/Helbing: Macroscopic Simulation of Widely Scattered
Synchronized Traffic States}
\tighten
\onecolumn
\twocolumn[\hsize\textwidth\columnwidth\hsize\csname @twocolumnfalse\endcsname
{\protect
\title{Macroscopic Simulation of Widely Scattered
Synchronized Traffic States}
\author{Martin Treiber$^1$ and Dirk Helbing$^{1,2}$}
\address{$^1$II. Institute of Theoretical Physics, University of Stuttgart,
         Pfaffenwaldring 57, D-70550 Stuttgart, Germany\\
$^2$Department of Biological Physics, E\"otv\"os University,
Budapest, Puskin u 5--7, H--1088 Hungary}

\date{\today}
\maketitle
\vfill
\begin{abstract}
Recently, a phase transition to synchronized congested traffic has been
observed in empirical highway data [B.~S.~Kerner and H.~Reh\-born, Phys. 
Rev. Lett. {\bf 79}, 4030 (1997)]. This hysteretic transition has been
described by a non-local, gas-kinetic-based traffic model
[D. Helbing and M. Treiber, Phys. Rev. Lett. {\bf 81}, 3042 (1998)] that,
however, did not display the wide scattering of synchronized states.
Here, it is shown that the latter can be reproduced by a mixture of different
vehicle types like cars and trucks. The simulation results are in good
agreement with Dutch highway data. 
\end{abstract}
} ]

%
Recent publications stressed the fact that, whereas in free traffic
the observed flow-density diagram is well described by a unique 
one-dimensional flow-density relation, in congested traffic the 
empirical data points are rather distributed over a 
two-dimensional region \cite{Kerner1,KernerSymp}.
In order to account for this fact, Krau\ss \cite{Krauss}
has recently proposed that
the driver behavior is changed in congested traffic 
compared to free traffic. Instead of this, Lenz {\em et al.}
\cite{Chr} and Schreckenberg \cite{Schreck}
have suggested that
the wide scattering of data is caused by an anticipation effect 
of drivers who not only react to the respective vehicle in front 
but also to the traffic dynamics further ahead. 
\par
In contrast to these "microscopic" approaches which 
simulate the interactions of individual vehicles, 
macroscopic models describe the evolution of
the macroscopic velocity
$V(x,t) = \erw{v_{\alpha}}$ and the vehicle density
$\rho(x,t) = \erw{1/s_{\alpha}}$, which are local averages
of the "microscopic" velocities $v_{\alpha}$ of the vehicles $\alpha$
and their center-to-center distances $s_{\alpha}$. 
All fluctuating quantities like individual velocity variations 
or distance distributions are eliminated. 
This means that, in deterministic macro-simulations, 
all self-organized structures (like stop-and-go waves or congested
traffic) are smooth. 
\par
Therefore, some researchers
believe that, while the wide scattering of
the congested flow-density data may be reproduced by
{\em microsopic} traffic models, {\em macroscopic} ones will fail for
principal reasons. However, motivated by the circumstance that
the scattering has been observed in aggregated rather than single-vehicle
data, we are confident that a macroscopic simulation of this effect 
should be possible. 
\par
In the following, 
we will show that the scattering can be explained by the fluctuations 
caused by a heterogeneous traffic population, which enter the 
macroscopic simulations {\it via} the boundary conditions. 
We will distinguish cars 
and trucks characterized by different sets of parameter values. These define 
two equilibrium flow-density relations of pure car traffic and pure truck
traffic, respectively, which are close to each other at small vehicle
densities, but considerably different in the congested density regime.
For mixed traffic, we interpolate between both parameter sets and, hence,
between both equilibrium relations, using a weighted average. The 
weights are extracted from real traffic data 
by determining the proportion of long vehicles ("trucks").
This method allows to simulate a uni-directional multi-lane freeway 
by an effective one-lane model for one car species with average, but varying 
parameter values. Our simulations 
are carried out with the empirically measured boundary conditions,
and the results are quite realistic. 
%
%
\par
For the simulations, we use
the macroscopic, gas-kinetic-based traffic model (GKT model)
\cite{HTLett,overview}, which shows
a realistic instability diagram and the characteristic properties of
traffic flows \cite{overview} demanded by Kerner and Reh\-born \cite{Kerner1}.
More importantly, this model is able to describe the hysteretic phase
transition to congested states with high traffic flow \cite{KernerLett}
(called "synchronized traffic" \cite{Kerner1})
which typically occurs behind on-ramps, gradients, or other bottlenecks
of busy freeways \cite{HTLett}.
\par
According to the gas-kinetic-based model, the evolution of the vehicle density
$\rho(x,t)$ in dependence of the time $t$ and the position $x$ 
along the freeway is given by the continuity equation
\begin{equation}
\label{eqrho} 
 \frac{\partial \rho}{\partial t} + 
 \frac{\partial (\rho V)}{\partial x} 
 = \frac{Q_{\rm rmp}}{nL} \, .
\end{equation}
Here, $V(x,t)$ denotes the average velocity of the vehicles. 
At on-ramps (or off-ramps), the source term $Q_{\rm rmp}/(nL)$
is given by the actually observed inflow $Q_{\rm rmp}>0$ 
from (or outflow $Q_{\rm rmp}<0$ 
to) the ramp, divided by
the merging length $L$ and by the number $n$ of lanes.
Otherwise it is zero, reflecting the conservation of the number of
vehicles. The average velocity obeys the equation
\begin{equation}
 \frac{\partial V}{\partial t}  
 + \!\!\!\!\underbrace{V \frac{\partial V}{\partial x}}_{\rm Transport Term}
 \!\!\!\! = \underbrace{- \frac{1}{\rho} \frac{\partial (\rho \theta)}
 {\partial x}}_{\rm Pressure Term}
 + \underbrace{\frac{1}{\tau} ( U - V )}_{\rm Relaxation Term} .
\label{geschwin}
\end{equation} 
According to this, the temporal change of the average velocity is given
by a transport term (caused by a propagation of the velocity profile
with $V$), a so-called pressure term (that reflects dispersion effects
due to the finite velocity variance $\theta$ of the vehicles), and 
a relaxation term (describing the adaptation to a dynamic equilibrium
velocity $U$ with a certain relaxation time $\tau$). In our gas-kinetic-based
model, the analytically derived
formula for the dynamical equilibrium velocity is
\be
\label{Vedyn}
 U
 = V_0\left[
     1 - \frac{\theta+\theta'}{2 A(\rhomax)}
          \left( \frac{\rho' T}{1-\rho'/\rhomax} \right)^2
          B(\delta_V) \right] \, , 
\ee
where $V_0$ is the desired (maximum) velocity, $T$ the average time headway at
large densities, and $\rho_{\rm max}$ the maximum vehicle density.
A prime indicates that the corresponding variable is taken at the 
advanced "interaction point" $x' = x + \gamma (1/\rho_{\rm max}
+ TV)$ rather than at the actual position $x$. This accounts for the
anisotropic anticipation behavior of drivers.
The monotonically increasing "Boltzmann factor" 
\begin{equation}  
\label{B}
B(\delta_V) =   2 \left[ 
    \delta_V \frac{\mbox{e}^{-\delta_V^2/2}}{\sqrt{2\pi}}
           + (1+\delta_V^2) 
    \int_{-\infty}^{\delta_V} dy \, \frac{\mbox{e}^{-y^2/2}}{\sqrt{2\pi}}
                 \right]
\end{equation}
can be derived from gas-kinetic formulas \cite{overview} and 
describes the dependence of the braking interaction on
the dimensionless velocity difference 
$\delta_V = (V-V')/\sqrt{\theta+\theta'}$. Finally, the dynamics of
the variance can be approximated by the constitutive relation
\begin{equation}
 \theta(x,t) =  \left[ A_0 + \Delta A \tanh 
      \left(\frac{\rho(x,t)-\rho_{\rm c}}{\Delta\rho} \right)
  \right] V^2(x,t) \, ,
\label{const}
\end{equation}
where the coefficients $A_0=0.008$, 
$\Delta A = 0.015$,
$\rho_{\rm c} = 0.28\rho_{\rm max}$,
and $\Delta\rho = 0.1\rho_{\rm max}$
have been obtained from single-vehicle data \cite{overview}. 
\par 
The velocity-density relation resulting for this model in
spatially homogeneous and stationary equilibrium reads
\be
V_{\rm e}(\rho) = \frac{\tilde{V}^2}{2V_0}
       \left( - 1 + \sqrt{1 + \frac{4 V_0^2}{\tilde{V}^2}}
       \right)
\ee
with
\be
\label{tilV}
\tilde{V}(\rho) = \frac{1}{T} \left(\frac{1}{\rho}-\frac{1}{\rho_{\rm max}}
 \right) \sqrt{\frac{A(\rho_{\rm max})}{A(\rho)} } \, .
\ee
This also determines the equilibrium traffic flow by
\be
\label{Qe}
Q_{\rm e}(\rho) = \rho V_{\rm e}(\rho) \, ,
\ee
which, for a given parameter set, is a one-dimensional curve. However,
as will be shown in the following, the empirically observed two-dimensional 
region of "synchronized" congested states can be reproduced by simulating
a mixture of different vehicle types. Although it has not been 
stressed clear enough, it is known from microsimulations
that heterogeneous traffic produces considerable fluctuations
of the aggregate quantities like the vehicle density and the 
average velocity \cite{two,zell,nat}. 
Nevertheless, we do not need to carry out microsimulations to
account for the two-dimensional scattering of synchronized traffic states.
It is 
sufficient to simulate traffic in a macroscopic way
with empirically obtained boundary conditions, including the varying
proportion of long vehicles ("trucks").
Thus, we do not need to assume other 
sources of fluctuations than observed ones. 
A reasonable agreement with empirical data can
already be reached by distinguishing two vehicles types only, short vehicles
("cars") and long ones ("trucks", with a length of at least 7 m). 
%
%
Each type is characterized by its own paramter set.
For the cars we assume
a desired velocity $V_0 = 112$\,km/h, an average time headway $T
=1.0$\,s at large densities, and a maximum density 
$\rhomax = 110$\,vehicles/km.
Trucks are described by the parameters 
$V_0 = 90$\,km/h, $T=5$\,s, and $\rhomax = 100$\,vehicles/km.
The remaining model parameters are the same for both types:
$\tau = 25$\,s and $\gamma = 1.6$. The parameters
in the constitutive relation (\ref{const}) for the variance have also been
chosen identical. 
\par
According to the philosophy of macroscopic models, we now define
time-dependent "effective" model parameters $X(t)$
as weighted averages of the respective car and truck parameters
$X_{\rm car}$ and $X_{\rm truck}$:
%
\be
\label{Xt}
X(t) = p_{\rm truck}(t) X_{\rm truck}
     + [1- p_{\rm truck}(t)] X_{\rm car}.
\ee
Here, 
$p_{\rm truck}(t)$ is the
proportion of trucks averaged over a time interval $\Delta t$ 
around $t$ [Figure~\ref{figBCtruck}(a)]. 
Although the approximations behind the resulting
"effective" macroscopic simulation 
model are rather crude, it yields a surprisingly good agreement
with empirical data. Even better results are expected for
macroscopic models which explicitly take into account different vehicle types
and lane-changing interactions among the freeway lanes \cite{multi}. 
\par
We simulated traffic flow on a section of the Dutch two-lane motorway A9 from
Haarlem to Amsterdam (Figure~\ref{sketch}) from the detector
cross-section D1 (0\,km) to D6 (5.7\,km). For this purpose,
the measured single-vehicle data were aggregated to 1-minute averages of
the velocity, traffic flow, and truck proportion (Figure~\ref{figBCtruck}).
Between 7:30 am (450 min) and 9:30 am (570 min) in the morning of
November 2, 1994, we find transitions from a low-density regime to a 
high-density regime corresponding to transitions between free and
congested traffic. Figure~\ref{figtime} 
illustrates that the congested state at D2 is connected
with a considerable velocity drop,
while the flow is decreased only by about 10\%, both in the 
empirical data and in the simulation. In addition,
the congested traffic state relaxes
to free traffic downstream of the on-ramps 
[Figures \ref{figrhoQ}(c) and (d)]. 
A comparison
with Figures~1(b) and 3(c) of Ref. \cite{KernerLett} 
suggests that the congestion in the investigated data
corresponds to synchronized traffic.
\par
As inflow and outflow boundary conditions, 
we used the data of the cross-sections
D1 and D6, respectively, as shown in Figure~\ref{figBCtruck}(b).
There are two on-ramps and one off-ramp in the considered section.
For all ramps, we use the empirical data of the traffic flow $Q_{\rm rmp}$,
divided by the number $n=2$ of lanes \cite{HTLett} 
[Figure~\ref{figBCtruck}(c)], and
assume a merging length $L=200$\,m.
\par
In the simulation, congested traffic first sets in at $t\approx 450$\,min 
near the on-ramp at D3a,
which agrees well with the empirical findings.
We started the simulation 50 min earlier to
eliminate any effects of initial conditions, and to show the
spontaneous nature of the transition [Figure~\ref{figtime}(a)].
In Figure~\ref{figrhoQ}(b), the free traffic flow before the breakdown
($t< 450$\,min) and after the recovery
($t>570$\,min) is delineated by the points at
the low-density (left) side of the diagram, which more or less define
a one-dimensional curve. In contrast,
the congested traffic state is represented by the points at the
high-density (right) side, which are distributed over a two-dimensional region.
Some minutes later, the front of the congested state crosses 
the on-ramp at D2, which causes congested traffic upstream
of it. In accordance with the mechanism
of the formation of synchronized traffic proposed in \cite{HTLett},
the congested state upstream of D2 has a lower flow 
and a higher density [Figure~\ref{figrhoQ}(a)] than that between D2
and D3a. The congested states are sustained for nearly two hours, until the
inflows from both the main road and the on-ramps are considerably decreased, 
which shows the hysteretic nature of the transition.
\par
%
%
Summarizing our results, one can say that the macroscopic, gas-kinetic-based
traffic model allows to simulate synchronized
traffic, including the associated scattering of the flow-density data
in the congested regime.
Simulations of this model with only one vehicle type
\cite{HTLett,numerics} suggested that the phenomenon of synchronized
traffic as such (i.e., high traffic flows at low velocities) 
does not depend on the existence of different types of vehicles.
However, as is often the case for self-organized non-chaotic
patterns resulting from deterministic dynamics,
the flow-density diagram is essentially one-dimensional.
\par
In this Letter, we showed that a realistic scattering in the flow-density plane
can be simulated by distinguishing several vehicle 
types with different parameter sets, the 
measured proportions of which are the weights 
for determining the time-dependent
``effective'' parameter set. A reasonable agreement with empirical data from
Dutch highways is already obtained for two different vehicle types, 
cars and trucks. Our results also
indicate that, when studying dynamical phenomena
in empirical traffic data, it is highly recommended
to thoroughly analyze the proportion of trucks, which shows
surprisingly large variations [Figure~\ref{figBCtruck}(a)].
\par
Notice that 
the assumed parameter variations due to a changing truck fraction
can explain both the
relatively low scattering of flow-density data in the low-density regime 
and the wide scattering in the regime of congested traffic.
While the small amount of scattering for free traffic at low densities
is caused mainly by variations of the individual desired velocity
(with a standard deviation of about 10\% of the 
mean value), the main reason for the considerable scattering 
for congested traffic at densities above 30\,vehicles/km
are the variations of the time headway (which are of the order
of 100\%). There are other effects that influence
scattering of congested traffic (e.g., lane changes), but they will only
increase the scattering of the flow-density data.
\par
Finally, we mention that the equations of the gas-kinetic-based 
traffic model together
with relation \refkl{Xt} for the stochastic quantities $V_{\rm e}$, $T$, and
$\rhomax$, represent stochastic partial equations with
multiplicative noise. 
They may serve as prototype for introducing
stochasticity into macroscopic equations in a controlled
and empirically
justified manner.


\subsection*{Acknowledgments}
The authors want to thank for financial support by the BMBF (research
project SANDY, grant No.~13N7092) and by the DFG (Heisenberg scholarship
He 2789/1-1). They are also grateful to Henk Taale and 
the Dutch {\it Ministry of Transport,
Public Works and Water Management} for supplying the freeway data.



\newpage

\begin{figure}[htbp]
\unitlength0.95cm
\begin{center}
\begin{picture}(9.22,1.95)
\thicklines
\put(1.33,1.8){\makebox(0,0){\footnotesize Rottepolderplein}}
\put(2.88,1.8){\makebox(0,0){\footnotesize S\,17\vphantom{p}}}
\put(7.56,1.8){\makebox(0,0){\footnotesize Badhoevedorp}}
\put(4.51,1.3){\vector(1,0){0.4}}
\put(4.51,1.1){\vector(1,0){0.4}}
%
\thinlines
\put(0,1.4){\line(1,0){9.22}}
\dashline{0.1}(0,1.2)(9.22,1.2)
\put(1,0.5){\line(0,1){0.9}}
\put(2.06,0.5){\line(0,1){0.9}}
\put(2.56,0.5){\line(0,1){0.9}}
\dashline{0.07}(3.01,0.5)(3.01,1.4)
\put(3.51,0.5){\line(0,1){0.9}}
\put(4.71,0.5){\line(0,1){0.9}}
\put(6.71,0.5){\line(0,1){0.9}}
\put(7.41,0.5){\line(0,1){0.9}}
\put(7.71,0.5){\line(0,1){0.9}}
\put(8.42,0.5){\line(0,1){0.9}}
%
\put(1,0.2){\makebox(0,0){\footnotesize D1}}
\put(1.96,0.2){\makebox(0,0){\footnotesize D2}}
\put(2.51,0.2){\makebox(0,0){\footnotesize D3}}
\put(3.06,0.2){\makebox(0,0){\footnotesize D3a}}
\put(3.61,0.2){\makebox(0,0){\footnotesize D4}}
\put(4.71,0.2){\makebox(0,0){\footnotesize D5}}
\put(6.71,0.2){\makebox(0,0){\footnotesize D6}}
\put(7.26,0.2){\makebox(0,0){\footnotesize D7}}
\put(7.86,0.2){\makebox(0,0){\footnotesize D8}}
\put(8.42,0.2){\makebox(0,0){\footnotesize D9}}
\put(0,1){\line(1,0){0.75}}
\put(0.75,1){\line(1,-1){0.15}}
\put(1,1){\line(1,-1){0.15}}
\put(1,1){\line(1,0){1.06}}
\put(2.06,1){\line(-1,-1){0.15}}
\put(2.31,1){\line(-1,-1){0.15}}
\put(2.31,1){\line(1,-1){0.15}}
\put(2.56,1){\line(1,-1){0.15}}
\put(2.56,1){\line(1,0){0.45}}
\put(3.01,1){\line(-1,-1){0.15}}
\put(3.26,1){\line(-1,-1){0.15}}
\put(3.26,1){\line(1,0){3.9}}
\put(7.16,1){\line(1,-1){0.15}}
\put(7.41,1){\line(1,-1){0.15}}
\put(7.41,1){\line(1,0){0.3}}
\put(7.71,1){\line(-1,-1){0.15}}
\put(7.96,1){\line(-1,-1){0.15}}
\put(7.96,1){\line(1,0){1.26}}
\end{picture}
\end{center}
\caption[]{Overview of the evaluated stretch of the Dutch Highway A9 from
Haarlem to Amsterdam.\label{sketch}}
\end{figure}
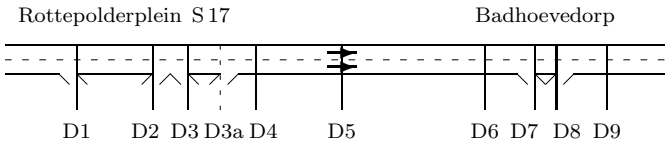
\begin{figure}

\vspace{0mm}

\begin{center}
   \includegraphics[width=77mm]{\pathfigs/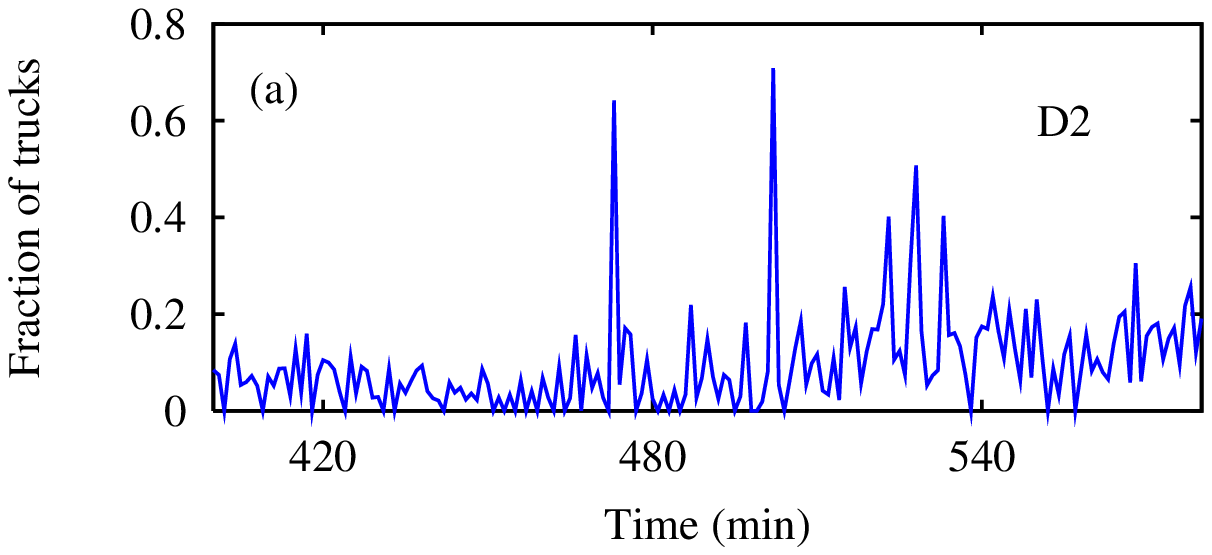} \\[0mm]
   \hspace*{-2mm}\includegraphics[width=80mm]{\pathfigs/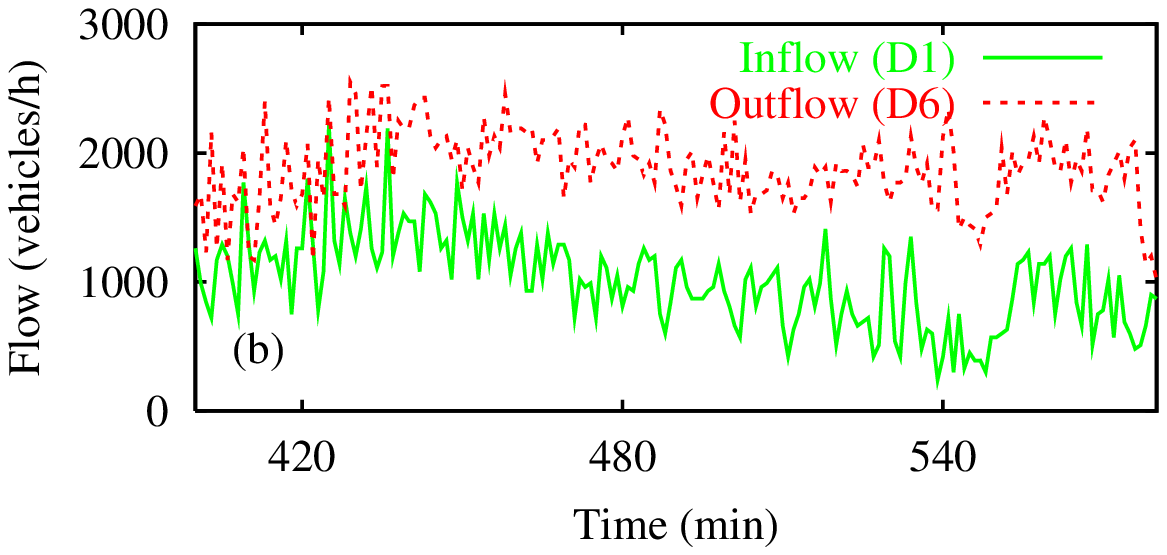} \\[0mm]
   \hspace*{-2mm}\includegraphics[width=80mm]{\pathfigs/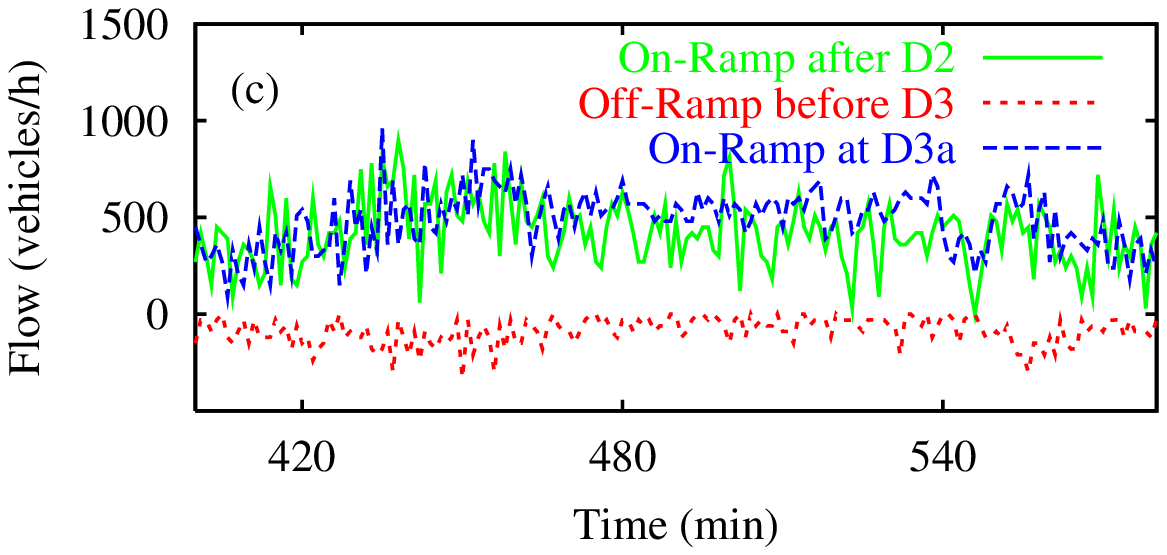} 
\end{center}
\caption{\label{figBCtruck}
(a) Proportion of trucks, from one-minute averages.
(b) Upstream and downstream boundary conditions for the
flow, taken from measured one-minute data at the cross-sections D1 and D6,
respectively.
(c) Flows of the three ramps in the considered section.
The off-ramp at
detector D1 was left out. It leads only to  changes of
the traffic situation upstream of the cross-section
D1 for which no data were available.} 
\end{figure}
\begin{figure}
\vspace{0mm}
\begin{center}
   \includegraphics[width=80mm]{\pathfigs/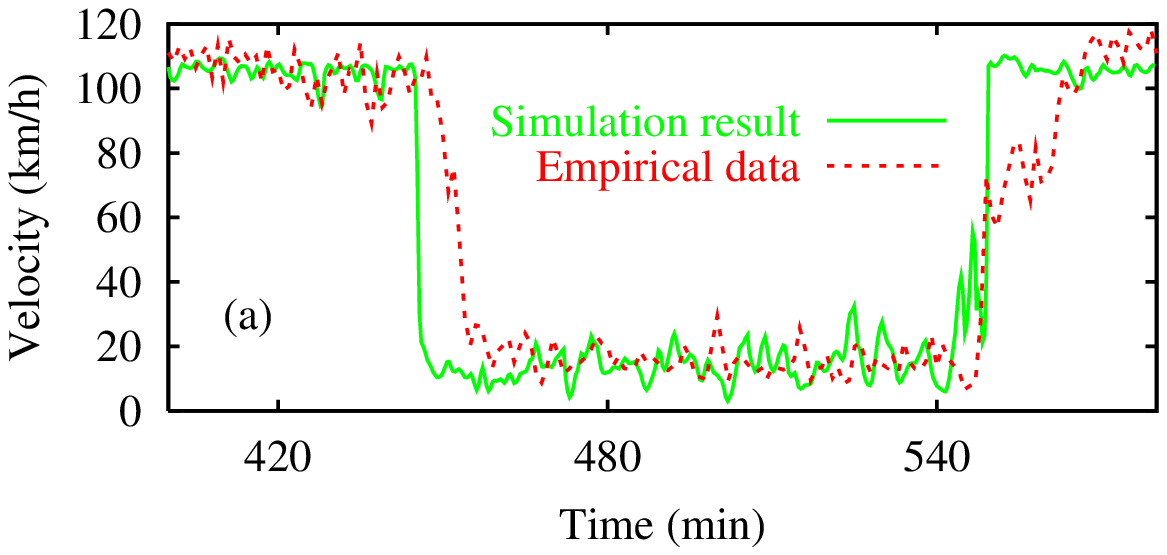} \\[0mm] 
   \includegraphics[width=80mm]{\pathfigs/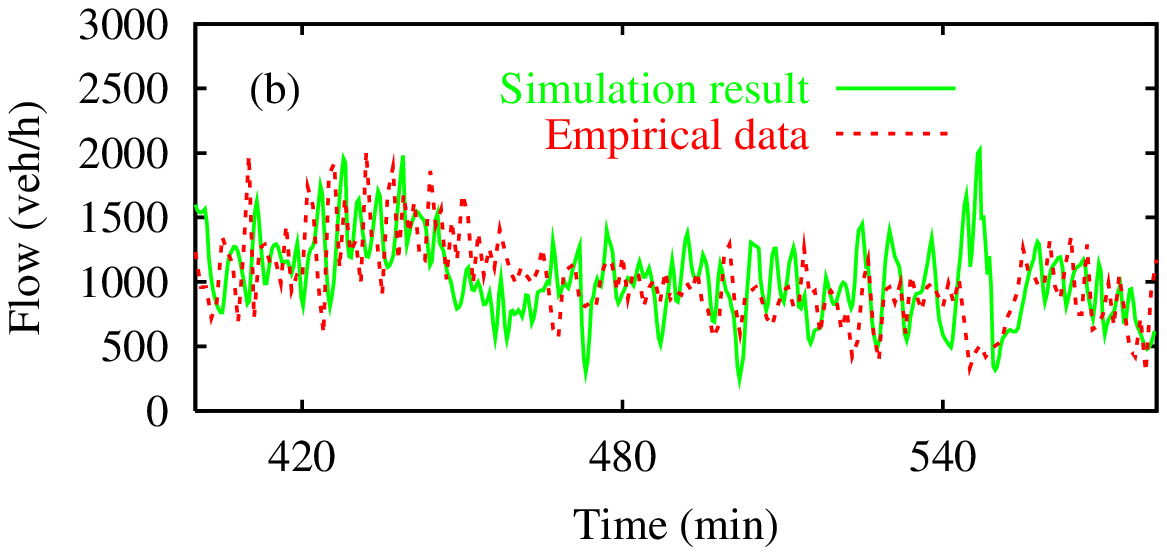}
\end{center}
\caption{\label{figtime}
(a) Velocity, and (b) traffic flow at D2 according to the model,
in comparison with the empirical one-minute data.
The breakdown of velocity is a result of a dynamical transition, since
neither the initial conditions, nor the
boundary conditions, or the ramp flows used in the simulations 
contain any significant peaks. 
}
\end{figure}
\begin{figure}

\vspace{0mm}

\begin{center}
   \includegraphics[width=44mm]{\pathfigs/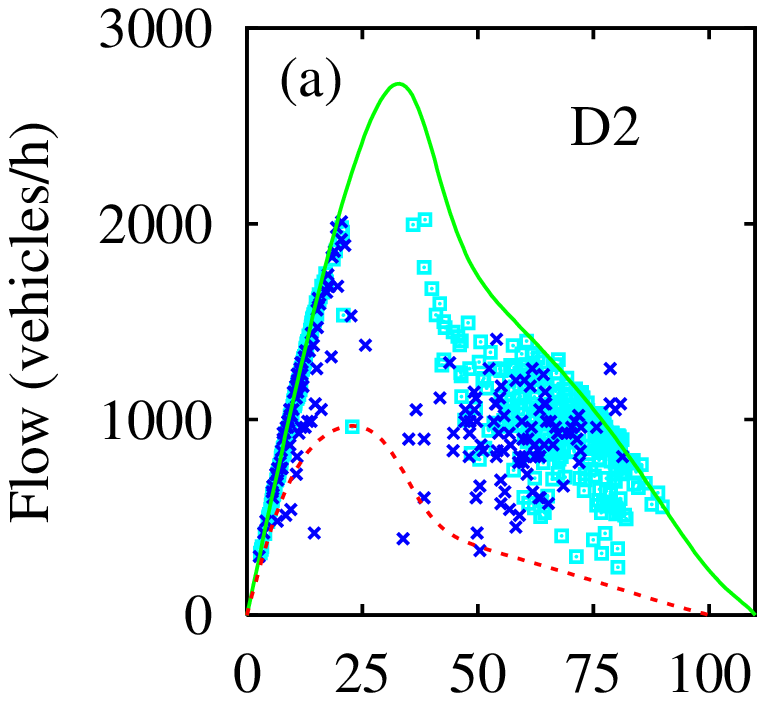} \hspace{-5mm}
   \includegraphics[width=40mm]{\pathfigs/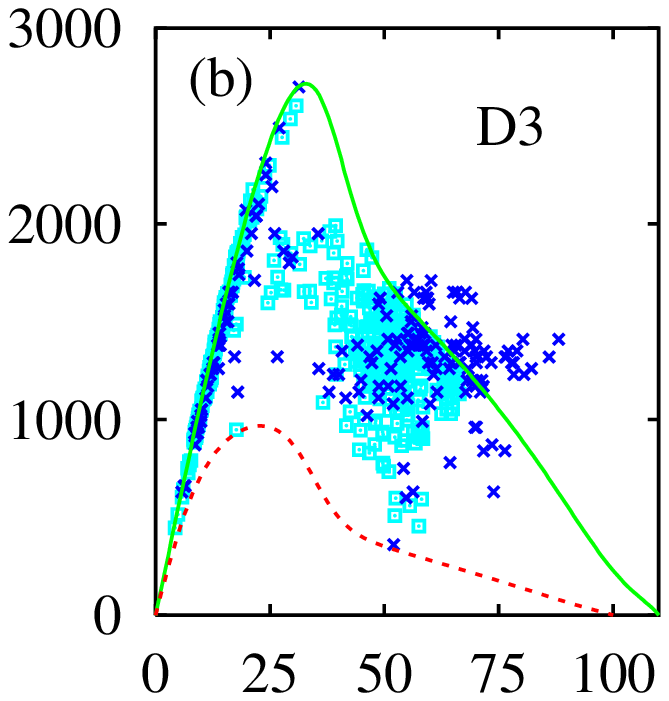} \\[0mm]
   \includegraphics[width=44mm]{\pathfigs/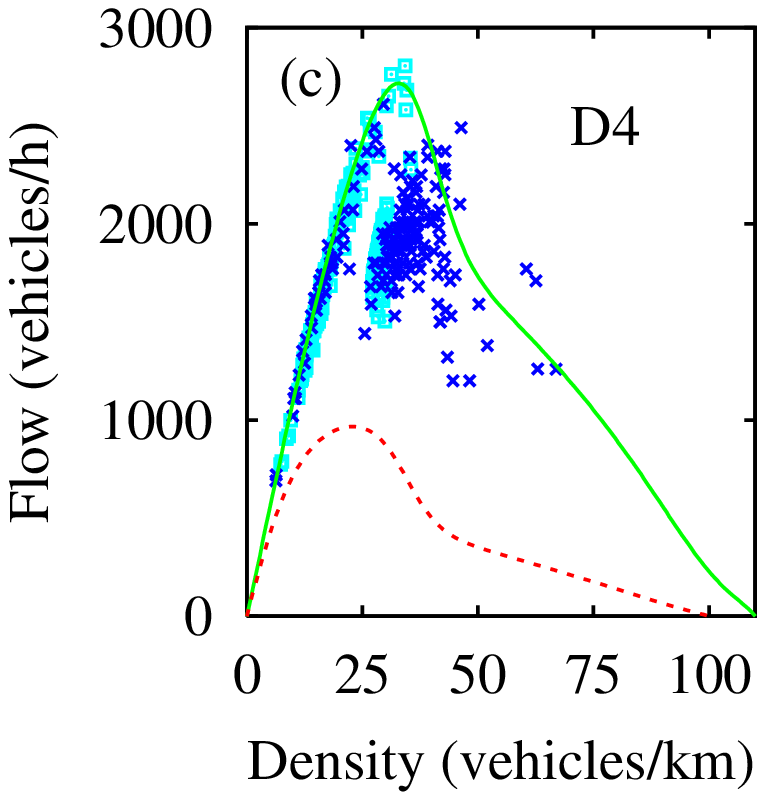} \hspace{-5mm}
   \includegraphics[width=40mm]{\pathfigs/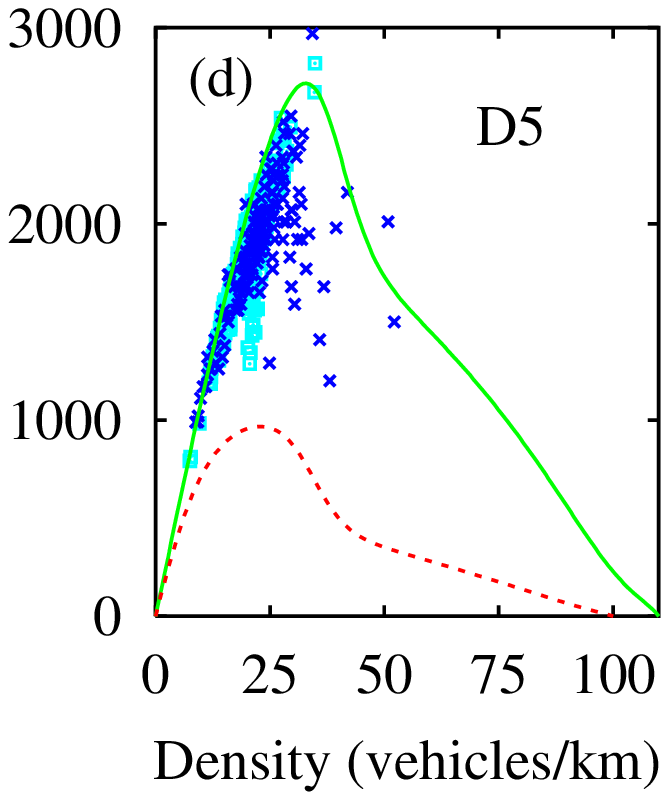} 

\end{center}
\caption{\label{figrhoQ}
The displayed points in density-flow space correspond
to empirical 1-minute data 
(dark crosses) and related simulation results (grey boxes), 
separately for the cross sections D2, D3, D4, and D5. 
The simulations manage to reproduce both the quasi-linear 
flow-density relation at small densities and the scattering over 
a two-dimensional region at high densities.
For comparison, we have displayed the equilibrium flow-density relations
for traffic consisting of 100\% cars (---),
and  100\% trucks (-~-~-).
}

\end{figure}



\begin{references}
\bibitem{Kerner1}
Kerner B S and Reh\-born H 1996 Phys. Rev. E {\bf 53},  R4275.  

\bibitem{KernerSymp}
Kerner B S 1998 in {\em Proc. Third Int. Symp. Highway Capacity}, Vol. II,
ed Rysgaard R (Road Directorate, Copenhagen, Denmark).

\bibitem{Krauss} Krau{\ss} S 1998 {\em Microscopic Modelling of Traffic Flow} 
(PhD thesis, DLR, Cologne), FB 98-08.

\bibitem{Chr} Lenz H, Wagner C K, and Sollacher R 1998 
European Physical Journal, in print. 

\bibitem{Schreck} Schreckenberg M 1998 To be published.

\bibitem{HTLett} Helbing D and Treiber M 1998 Phys. Rev. Lett. {\bf 81},
3042.

\bibitem{overview}
Treiber M, Hennecke A, and Helbing D 1999
Phys. Rev. E {\bf 59}, in print.

\bibitem{KernerLett}
Kerner B S and Reh\-born H 1997 Phys. Rev. Lett. {\bf 49},  4030.

\bibitem{two} Nagel K, Wolf D E, Wagner P, and Simon P 1998
{Phys. Rev. E} {\bf 58}, 1425. 

\bibitem{zell} Helbing D and Schreckenberg M
1998 ``Cellular automata simulating experimental
properties of traffic flows'', Phys. Rev., submitted. 

\bibitem{nat} Helbing D and Huberman B A 1998 ``Moving like a solid block'',
Nature, in print.

\bibitem{multi} Shvetsov V and Helbing D 1998 ``Macroscopic dynamics
of multi-lane traffic'', Phys. Rev. E, submitted.  

\bibitem{numerics} Helbing D and Treiber M 1999 
  ``Numerical simulation of macroscopic traffic equations'', 
  Computing in Science and Engineering, in print.
\end{references}
\end{document}